\documentclass[sn-nature]{sn-jnl}

\usepackage{upgreek}
\usepackage{graphicx}%
\usepackage{multirow}%
\usepackage{amsmath,amssymb,amsfonts}%
\usepackage{amsthm}%
\usepackage{mathrsfs}%
\usepackage[title]{appendix}%
\usepackage{xcolor}%
\usepackage{textcomp}%
\usepackage{manyfoot}%
\usepackage{booktabs}%
\usepackage{algorithm}%
\usepackage{algorithmicx}%
\usepackage{algpseudocode}%
\usepackage{listings}%
\usepackage[utf8]{inputenc}
\DeclareUnicodeCharacter{00D7}{$\times$}
\DeclareUnicodeCharacter{2009}{\hspace{0.166em}}
\DeclareUnicodeCharacter{03BB}{$\uplambda$}
\usepackage{hyperref}

\theoremstyle{thmstyleone}%
\theoremstyle{thmstyletwo}%

\theoremstyle{thmstylethree}%

\begin{document}

\title[Non-uniform programmable photonic waveguide meshes]{Non-uniform programmable photonic waveguide meshes}


\author*[1]{\fnm{Cristina} \sur{Catalá-Lahoz}}\email{ccatala@iteam.upv.es}
\author*[1]{\fnm{Jose} \sur{Capmany}}\email{jcapmany@iteam.upv.es}

\affil*[1]{\orgdiv{Photonics Research Labs, iTEAM Research Institute}, \orgname{Universitat Politècnica de València}, \orgaddress{\street{Camí de vera s/n}, \city{Valencia}, \postcode{46022}, \state{Valencia}, \country{Spain}}}


\abstract{Programmable integrated photonics has emerged as a powerful platform for implementing diverse optical functions on a single chip through software-driven reconfiguration. At the core of these processors, photonic waveguide meshes enable flexible light routing and manipulation. However, recirculating waveguide meshes are fundamentally limited by the fixed dimensions of their unit cells, constraining their spectral and temporal resolution. These limitations hinder broadband signal processing and high-precision delay-line applications.  Here, we introduce the concept of non-uniform programmable waveguide meshes by incorporating defect cells into an otherwise uniform hexagonal architecture. These defect cells preserve an external hexagonal perimeter for seamless integration while embedding smaller internal subcells that modify the spectral response via the Vernier effect. By coupling cells with different optical path lengths, we achieve a free spectral range multiplication of up to tenfold, reaching $133$~GHz—surpassing the capabilities of conventional silicon ring resonators. Additionally, we demonstrate a dramatic reduction in sampling times from $75$~ps to $7.5$~ps, enabling ultra-fast optical signal processing and high-precision delay tuning. This approach unlocks new degrees of freedom in programmable photonic circuits, offering enhanced spectral and temporal tunability. Beyond broadband signal processing, it paves the way for advanced applications in topological photonics, quantum information processing, and high-speed optical computing.}

\keywords{Programmable photonics, Non-uniform mesh, Vernier effect, Silicon photonics}



\maketitle

\section{Introduction}
Programmable integrated photonics (PIP) \cite{Capmany2020, Bogaerts2020} is a new technology paradigm that aims at designing common integrated optical hardware resource configurations, capable of implementing an unconstrained variety of functionalities by suitable programming. The core of a PIP processor is a photonic waveguide mesh \cite{Zhuang2015, Perez2016}, a two-dimensional lattice that provides a regular and periodic geometry, formed by replicating unit cells. The mesh connectivity determines the possible functions of the programmable circuit and how it can be configured. Broadly, waveguide meshes can be classified into (1) \textit{forward-only} \cite{Harris2018, Miller2013, Clements2016, Reck1994, Ribeiro2016}, where the light flows from one side of the mesh to the other, and (2) \textit{recirculating} \cite{Zhuang2015, Perez2016, Perez2017, Perez-Lopez:19, Dai2024}, where light can also be routed in loops and even back to the input ports. 

Recirculating meshes typically adopt hexagonal \cite{Perez2016}, rectangular \cite{Zhuang2015}, or triangular \cite{Perez-Lopez:19} geometries, where each side of the cell is formed by a tunable basic unit (TBU) composed of two integrated waveguides connected by means of a balanced Mach–Zehnder interferometer (MZI) or a gate \cite{Perez2016, Miller2015, Wang2020, Perez2018, Rahim2018}. The mesh is then programmed to manipulate the flow of light on a chip by electrically controlling the set of tunable analogue gates \cite{Perez2020, Gao2024, Miller:22} as crossbar switches or as variable couplers with independent power division ratio and phase shift. Light is then distributed and spatially rerouted to implement various linear functions by interfering signals along different paths. PIP circuits enable broadband analogue processing, which finds applications in two relevant scenarios. The first includes applications where analogue processing is required, such as artificial intelligence \cite{Tait2014,Shastri2021}, deep learning \cite{Shen2017}, quantum information systems \cite{Carolan2015, Harris2017, Wang2018}, and topological photonics \cite{On2024,Dai2024,Capmany2024}. The second includes applications, such as 5 or 6G telecommunications \cite{Perez2024}, signal processing \cite{Hong2025, Zhou2024, Najjar2024}, switching and data centre interconnections \cite{Quack2020,Hu2025,Ding2016}, hardware acceleration \cite{Feldmann2021}, and sensors \cite{Hu2017}, where photonic analogue processing can work in conjunction with high-speed digital electronic systems to overcome the expected limitations that arise from the death of Moore's law. 

However, the performance of recirculating waveguide meshes is inherently constrained by the fixed physical dimensions of their unit cells. A typical state-of-the-art cell in silicon photonics has a side length $\text{L}_{\text{s}}$ of around 450 to 900~$\upmu$m \cite{Capmany2020, Bogaerts2020} providing a propagation delay $\text{T}_{\text{s}}$ in the range of 6.3 to 12.6~ps. These values are fixed once the mesh is fabricated and limit the flexibility in the spectral and time domain responses of the circuits that can be implemented. For instance, in hexagonal meshes, minimum length cavities $\text{L}=6\text{L}_{\text{s}}$ provide round trips between $37.6$ and $75.2$~ps that result in a spectral period in the range of $13.3-26.6$~GHz, which is not enough for broadband applications requiring at least $40$~GHz or more. Similarly, the time sampling rate achievable in the meshes is not enough for applications requiring delay precisions in the range of less than $10$~ps. 

Here, we propose and experimentally demonstrate a solution to the above limitations that consists of embedding what we call, in analogy to solid-state physics, \textit{defect} cells \cite{Ashcroft2003,Yablonovitch1987,John1987} into the otherwise uniform 2D hexagonal mesh. The defect cells have an external hexagonal perimeter to fit into the regular mesh but are internally divided into subcells with a smaller perimeter. In this way, the individual cavities of the uniform and defect cells have slightly dissimilar spectral periods or free spectral range ($\text{FSR}_{\text{u}}$ and $\text{FSR}_{\text{d}}$) values and the mesh spectral limitation can be solved exploiting the Vernier effect \cite{Bogaerts2012,Hulme2013}. In particular, we experimentally show multiplication factors of 3.6 ($\text{FSR}=48$~GHz), 6 ($\text{FSR}=80$~GHz) and 10 ($\text{FSR}=133$~GHz) in a hexagonal waveguide mesh implemented on a Silicon on Insulator (SOI) chip, which exceeds those reported in previous publications using silicon-based ring resonators \cite{Zhuang2013}. In addition, we show reductions of sampling times from $75$~ps down to $7.5$~ps. The foreseeable reduction of cell sidelengths to values below $100$~$\upmu$m needed for scaling up programmable optical circuits \cite{Bogaerts2023} can open the path in combination with defect cells to the implementation of cavities offering spectral periods well beyond $1$~THz and sampling times in the femtosecond range. The insertion of these defect cells enables the implementation of \textit{non-uniform} programmable waveguide meshes where both individual and clusters of defect cells are incorporated into the uniform 2D layouts. This new type of structure may find applications in the implementation of novel topological photonic layouts.

\section{Non-uniform mesh}
\begin{figure}[h]
\centering
\includegraphics[width=\textwidth]{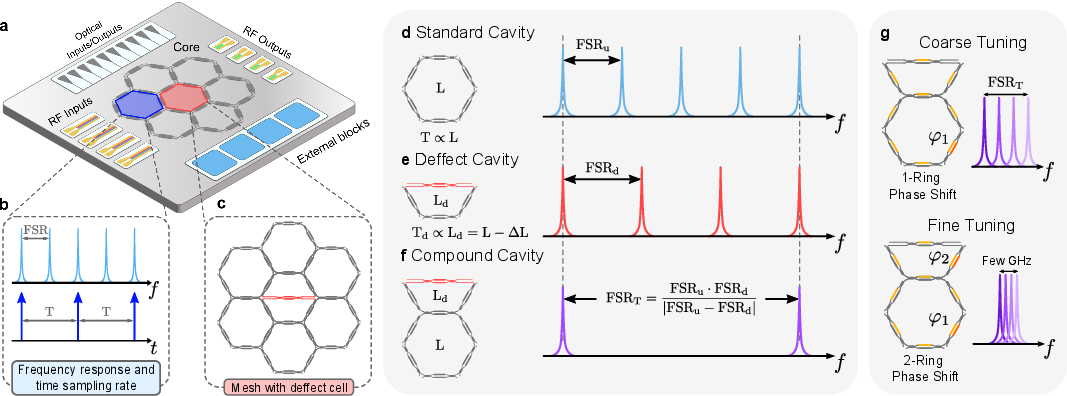}
\caption{Non-uniform programmable photonic meshes enabled by defect cells. (a) Schematic of a programmable photonic waveguide processor, comprising a reconfigurable core, optical and RF interfaces, and external control elements. (b) In conventional hexagonal meshes, cavities formed by six TBUs impose limitations on the FSR and temporal resolution. (c) A defect cell, created by introducing an additional TBU within a hexagonal cavity, alters the uniform structure, enabling new coupling mechanisms. (d, e) The standard and defect cavities exhibit different FSRs due to their distinct round-trip lengths ($\text{L}$ and $\text{L}_{\text{d}}$). (f) When combined, their spectral overlap produces an extended FSR via the Vernier effect, enhancing spectral tunability and enabling finer time resolution. (g) Coarse spectral tuning, controlled by adjusting a single cavity’s phase ($\varphi_1$), allows discrete spectral shifts, while fine spectral tuning, achieved by modifying both cavity phases ($\varphi_1$, $\varphi_2$), provides continuous adjustments for precise frequency control.}\label{fig1}
\end{figure}

A programmable photonic waveguide processor (Fig.~\hyperref[fig1]{1a}) consists of a programmable core, optical and radiofrequency input/output interfaces, and auxiliary components that enhance performance \cite{Perez2018, Rahim2018}. At the heart of this architecture lies a two-dimensional periodic waveguide mesh, typically structured in hexagonal, triangular, or square unit cells. In a hexagonal configuration, the fundamental cell perimeter—comprising six TBUs—determines the resonant cavity lengths. This inherent geometry imposes constraints on the free spectral range (FSR) and time sampling rate, limiting the processor's applicability in broadband and high-speed operations (Fig.~\hyperref[fig1]{1b}).

To overcome these limitations, we propose introducing defect cells within the regular mesh, creating a non-uniform structure that enables spectral period expansion by coupling cavities of different lengths and exploiting the Vernier effect (see Supplementary Note 1). As an example, Fig.~\hyperref[fig1]{1c} illustrates the insertion of an additional TBU to split one of the hexagonal cavities, forming a coupled system between a standard cavity with a round-trip length $\text{L}$ (Fig.~\hyperref[fig1]{1d}) and a defect cavity of length $\text{L}_{\text{d}}$ (Fig.~\hyperref[fig1]{1e}). The resulting compound cavity (Fig.~\hyperref[fig1]{1f}) exhibits an extended FSR $\text{FSR}\text{T}$ and a reduced time sampling rate due to the interplay between the original and defect cavity resonances. Alternative strategies for incorporating defect cells in hexagonal meshes are explored in detail in Supplementary Note 2.

Each TBU within the programmable mesh consists of a MZI with two phase shifters, one per arm, enabling precise control of coupling and phase at the output. In applications requiring only coupling adjustments, driving a single phase shifter is sufficient. However, for independent control of both coupling and phase, both phase shifters must be modulated. The coupling constant is given by $\kappa = \cos^2\bigl(\frac{\phi_2 - \phi_1}{2}\bigr)$, while the overall phase shift follows $\varphi = (\phi_1 + \phi_2)/2$.

Leveraging the reconfigurability of the mesh, we implement frequency-tunable ring resonators by dynamically adjusting the coupling and phase of selected MZIs. This flexibility extends to coupled cavities, where the Vernier effect allows for dual tuning mechanisms: coarse and fine tuning (Fig.~\hyperref[fig1]{1g}). Coarse tuning enables discrete frequency shifts across the extended FSR by modifying the phase $\varphi_1$ in a single cavity, whereas fine tuning provides continuous adjustments by simultaneously controlling $\varphi_1$ and $\varphi_2$. The two approaches differ in granularity: coarse tuning operates in discrete steps determined by $\text{FSR}_\text{u}$ or $\text{FSR}_\text{d}$, depending on the adjusted cavity, while fine tuning allows for precise spectral shifts down to a few GHz, governed by the resolution of the phase actuators. By combining both tuning strategies, the entire extended FSR can be covered with high spectral precision and without additional power overhead.

\section{Implementation}

\begin{figure}[h]
\centering
\includegraphics[width=\textwidth]{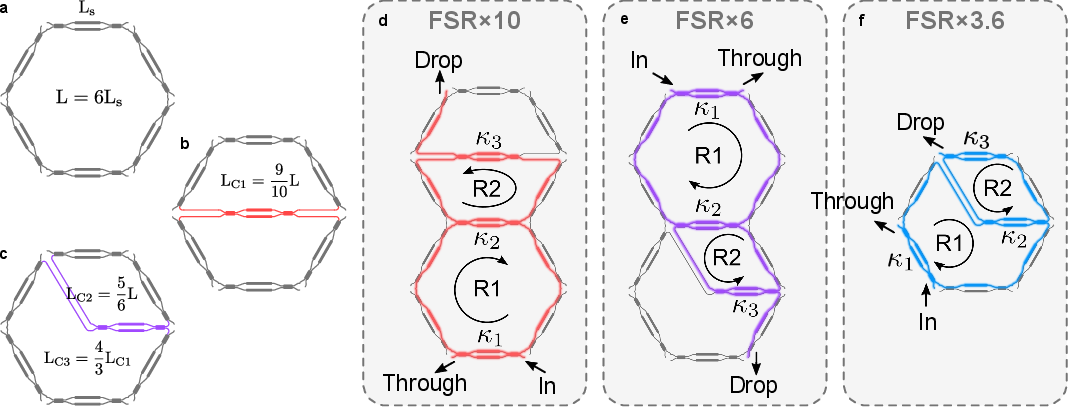}
\caption{Schematic representation of the hexagonal waveguide mesh. (a) Uniform basic cell composed of six tunable MZIs. (b) Defect cells created by splitting the hexagon into two equal parts with four MZIs each. (c) Alternative defect structure with asymmetric division into three and five MZI configurations. (d-f) Circuit diagrams of coupled double-ring resonators enhancing FSR.}
\label{fig2}
\end{figure}

We designed and implemented a programmable hexagonal waveguide mesh comprising three distinct types of cells, as illustrated in Fig.~\ref{fig2} (left). 

The basic cell (Fig.~\hyperref[fig2]{2a}) features six tunable MZIs, each with an optical path length of $\text{L}_\text{s}$, resulting in a total roundtrip length of $\text{L} = 6\text{L}_\text{s}$. To introduce defects, we defined two alternative cell structures. In the first case (Fig.~\hyperref[fig2]{2b}), the hexagon is bisected into two equal defect cells, each comprising four MZIs with length $\text{L}_{\text{C}1}$. In the second case (Fig.~\hyperref[fig2]{2c}), the hexagon is divided asymmetrically into cells with three and five MZIs, featuring lengths $\text{L}_{\text{C}2}$ and $\text{L}_{\text{C}3}$, respectively. The latter design maintains symmetry by incorporating an additional length on one side, ensuring that one MZI functions as a coupler within the ring. Importantly, both defect configurations preserve the external hexagonal perimeter, enabling seamless integration into a regular waveguide mesh.

The introduction of defect cells with varying cavity lengths facilitates an increase in the FSR through a serially coupled double-ring resonator. The FSRs of the coupled cavities must satisfy the condition $\text{FSR}_\text{T} = m_1 \text{FSR}_1 = m_2 \text{FSR}_2$, where $m_1$ and $m_2$ are natural, co-prime numbers \cite{rabus2007}. By coupling cavities of lengths $\text{L}_{\text{C}1}$ and $\text{L}_{\text{C}2}$, we achieve FSR enhancements of 10 and 6 times, respectively, compared to the uniform hexagonal cell. Additionally, coupling $\text{L}_{\text{C}2}$ and $\text{L}_{\text{C}3}$ within a single hexagonal cell further increases the FSR of $\text{L}_{\text{C}2}$ by a factor of 3, leading to an overall enhancement of 3.6 times relative to the hexagonal cavity. Figure~\hyperref[fig2]{2d-f} illustrate circuit diagrams demonstrating these coupled double-ring resonator structures. The transfer function at the drop port of the resulting resonator can be expressed as:

\begin{equation}
    \text{H}_{\text{Drop}} = \frac{\kappa_1 \kappa_2 \kappa_3 \alpha_\text{R1} \alpha_\text{R2} e^{j\frac{\omega \tau_\text{R1}}{2}} e^{j\frac{\omega \tau_\text{R2}}{2}}}{1 - t_1 t_2 \alpha_\text{R1}^2 e^{j\omega \tau_\text{R1}} - t_2 t_3 \alpha_\text{R2}^2 e^{j\omega \tau_\text{R2}}  + t_1 t_3 \alpha_\text{R1}^2 \alpha_\text{R2}^2 e^{j\omega \tau_\text{R1}} e^{j\omega \tau_\text{R2}}  }, \label{eq1}
\end{equation}

\noindent where $\kappa_{1,2,3}$ and $t_{1,2,3}$ represent the coupling and transmission coefficients of the tunable MZIs, $\tau_\text{R1}$ and $\tau_\text{R2}$ are the roundtrip delays, $\omega$ is the optical angular frequency, and $\alpha_\text{R1}$ and $\alpha_\text{R2}$ are the half-roundtrip loss coefficients. To maximize the extinction ratio of the Vernier-enhanced resonances, an optimization routine determines the optimal coupling parameters based on total ring losses.

A non-uniform four-cell hexagonal waveguide mesh was designed and fabricated as a photonic integrated circuit (PIC) using the silicon on insulator platform from Advanced Micro Foundry (see “Methods” for details). The foundry specifications include a waveguide loss of 1.1~dB/cm, multimode interferometer (MMI) insertion losses of 0.15~dB, and edge coupler insertion losses of 1.3~dB, all at a wavelength of 1550~nm. 

\begin{figure}[h]
\centering
\includegraphics[width=\textwidth]{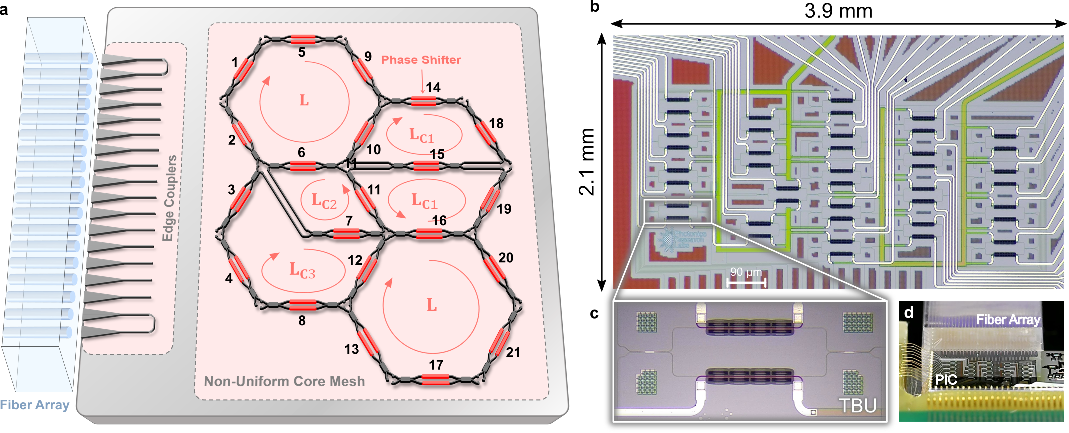}
\caption{Fabrication and packaging of the photonic circuit. (a) Conceptual layout of the PIC design. (b) Micrograph of the fabricated chip, showing the hexagonal waveguide mesh. (c) Low-power phase shifters utilizing suspended waveguides and trenches to minimize power consumption. (d) Packaged chip wirebonded to a PCB with fiber array alignment for optical coupling.}
\label{fig3}
\end{figure}

The fabricated chip (Fig.~\hyperref[fig3]{3b}) follows a flattened hexagonal mesh configuration (see Supplementary Note 3) with hexagonal sides of 900~$\upmu$m, including interconnecting waveguides for the 21 integrated MZIs. To reduce power consumption, the phase shifters employ suspended waveguides with trenches, achieving an efficiency of 1.26~mW/$\uppi$ and a tuning speed of ~3~kHz (Fig.~\hyperref[fig3]{3c}). The chip was electrically packaged onto a printed circuit board (PCB) for precise phase control, with 24 edge couplers for optical input/output interfacing. Alignment loops facilitate fiber array alignment during coupling (Fig.~\hyperref[fig3]{3d}).

\section{Experimental results}
We programmed three types of double-coupled ring Vernier resonators using the fabricated programmable mesh. To configure a photonic circuit within the mesh, we first characterized each TBU. Due to fabrication deviations, the TBUs are not perfectly balanced (see Supplementary Note 4). As a result, each TBU requires different current values depending on the desired coupling, which can be programmed via software using multichannel sources.

\begin{figure}[h] 
\centering 
\includegraphics[width=\textwidth]{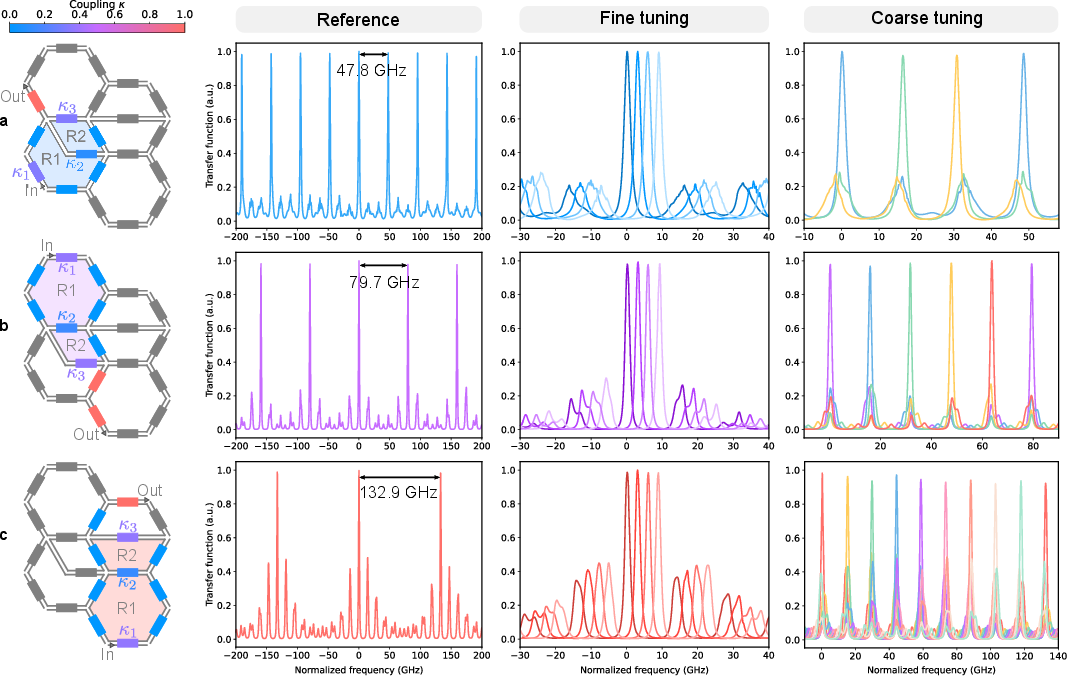} 
\caption{Experimental spectral results for three types of Vernier cavities formed by combinations of three different cells ($\text{L}_{\text{C}1}$, $\text{L}_{\text{C}2}$, and $\text{L}_{\text{C}3}$). (a) Coupling $\text{L}_{\text{C}2}$ and $\text{L}_{\text{C}3}$ results in a $3.6\times$FSR extension to $47.8$~GHz. (b) Coupling a hexagonal cavity with $\text{L}_{\text{C}3}$ results in a $6\times$FSR increase, reaching $79.7$~GHz. (c) Coupling a hexagonal cavity with $\text{L}_{\text{C}1}$ results in a $10\times$FSR increase, yielding $132.9$~GHz.}
\label{fig4} 
\end{figure}

For optical measurements, we used a tunable laser combined with a component tester to characterize the ring responses (see Supplementary Note 5 for setup details). Figure~\ref{fig4} shows experimental spectral results for the three types of Vernier cavities formed by combinations of three different cells: $\text{L}_{\text{C}1}$, $\text{L}_{\text{C}2}$, and $\text{L}_{\text{C}3}$. The reference spectrum column illustrates the spectral response of each circuit extended over multiple spectral periods. Coupling the $\text{L}_{\text{C}2}$ and $\text{L}_{\text{C}3}$ cavities (Fig.~\hyperref[fig4]{4a}) extended the FSR by a factor of 3.6 compared to a hexagonal cavity (see Supplementary Note 6), yielding an FSR of $47.8$~GHz. Similarly, coupling a hexagonal cavity with the $\text{L}_{\text{C}3}$ cavity increased the FSR by a factor of six, reaching $79.7$GHz (Fig.~\hyperref[fig4]{4b}). Finally, coupling a hexagonal cavity with $\text{L}_{\text{C}1}$ yielded a tenfold FSR increase, resulting in $132.9$GHz (Fig.~\hyperref[fig4]{4c}).

The reference spectrum can be tuned in two ways. Fine tuning is performed by operating one of the MZIs in each ring as a phase shifter, applying the same current to both rings. In our experiments, we adjusted the response in $3.2$~GHz steps, corresponding to a power difference of $\Delta \text{P} = 0.35$~mW per phase shifter. The maximum frequency shift is limited by the breakdown current or voltage of the phase shifter ($8.39$~mA).

Coarse spectral tuning is achieved by applying a phase shift to only one ring, altering its resonance while leaving the second ring fixed, thus adjusting the Vernier effect to match the next resonance of the second cavity. In Fig.~\hyperref[fig4]{4a,b}, coarse tuning was performed by shifting the response of the $\text{L}_{\text{C}3}$ and hexagonal rings, achieving a $15.95$GHz frequency step in both cases. For the circuit in Fig.~\hyperref[fig4]{4c}, coarse tuning of the hexagonal ring resulted in a $14.76$~GHz frequency step, enabling tuning across the entire FSR of $132.9$~GHz.

By combining fine and coarse tuning, we can continuously shift the response of these rings over the extended FSR, making the system highly reconfigurable in frequency. Table~\ref{tab1} lists the coupling parameters and current values used for each case, including those for both fine and coarse tuning.

The total losses per MZI are approximately $0.48$~dB, implying a loss factor of $\alpha = 0.85$ for a ring formed with a hexagonal cell. Consequently, the Vernier effect response deviates from the ideal case, where secondary resonances exhibit a lower extinction ratio compared to the primary coupled resonance. Reducing TBU losses should improve these responses, as discussed in the following section.

\begin{figure}[h] \centering \includegraphics[width=\textwidth]{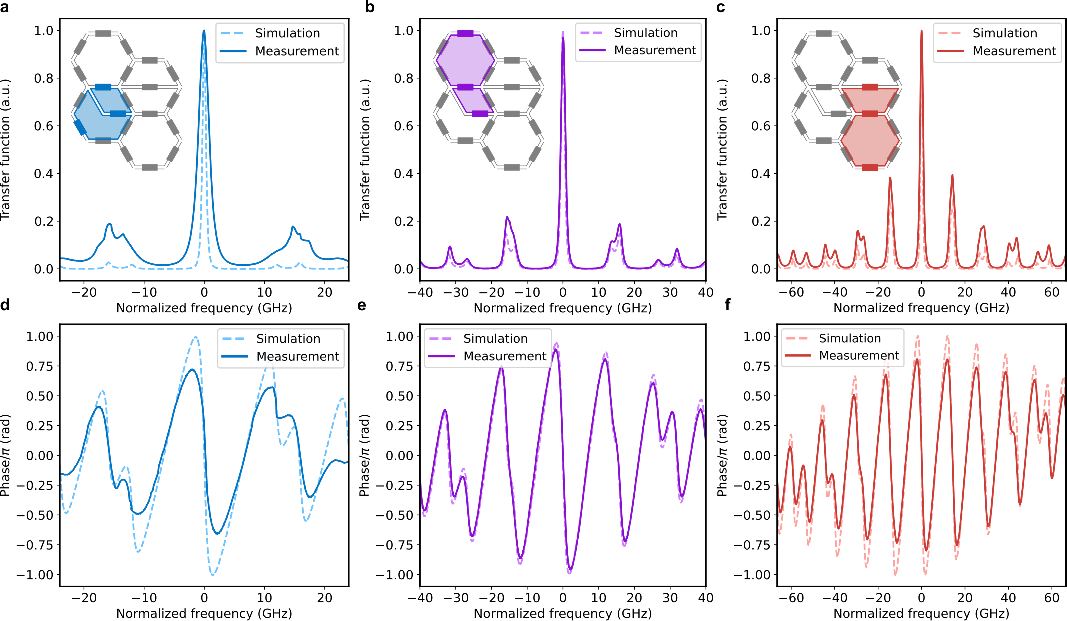} \caption{(a-c) Measured amplitude response for the three coupled cavity configurations. Experimental results (solid lines) are compared with the simulated response (dashed lines). The coupling and loss parameters used in the simulations are listed in Table~\ref{tab1}. (d-f) Corresponding phase response, derived using Kramers–Kronig relations, showing the interaction between the resonances of the two cavities. The primary resonance exhibits a $2\pi$ phase shift, while secondary resonances show a smaller phase variation.}\label{fig5} \end{figure}

To analyze this deviation, we compared the measured amplitude and phase responses with simulated results (1), using the coupling and loss parameters specified in Table~\ref{tab1}. Figure~\hyperref[fig5]{5a-c} show the amplitude responses for the three coupled cavity cases, where experimental measurements align closely with the simulated results. Figure~\hyperref[fig5]{5d-f} illustrate the corresponding phase responses, derived from the Kramers–Kronig relations (see Supplementary Note 7), demonstrating the interaction between the resonances of both cavities. The primary resonance exhibits a $2\pi$ phase shift, while the secondary resonances show a smaller, more gradual phase variation.

\begin{figure}[h] 
\centering 
\includegraphics[width=\textwidth]{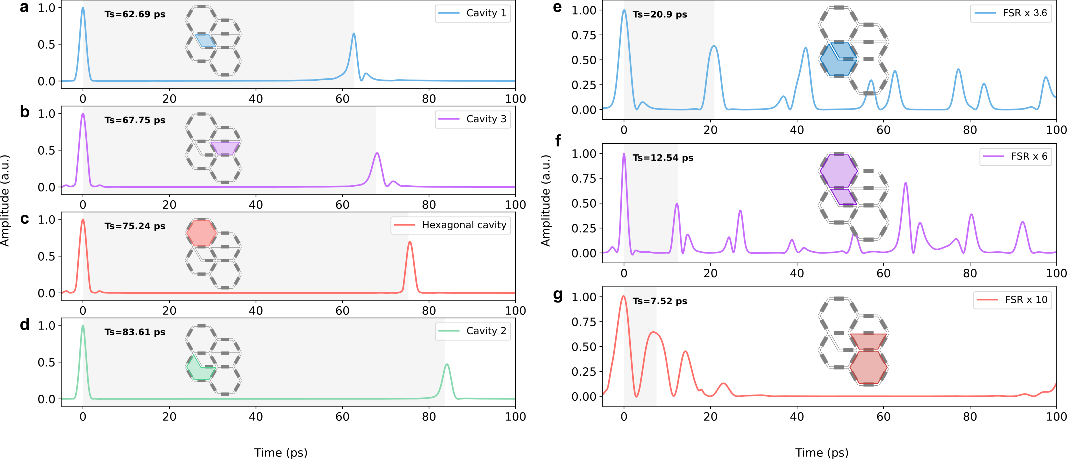} 
\caption{(a-d) Temporal responses for each independent cavity measured using 1-picosecond Gaussian pulses, with resolutions ranging from $62.69$ to $83.61$~ps. (e-f) Temporal responses of the coupled cavities of different lengths, demonstrating a significant reduction in sampling times to a range of $7.52$ to $20.9$~ps. This improvement is particularly useful for applications requiring precise delay measurements.}
\label{fig6} 
\end{figure}

Additionally, inserting defect cells allows for a reduction in sampling time resolution, addressing the limitation found in uniform waveguide meshes. To demonstrate this, we measured the temporal response of each cavity within the mesh using 1-picosecond Gaussian pulses. Figure~\hyperref[fig6]{6a-d} show the sampling time between consecutive pulses for each independent cavity, achieving resolutions ranging from $62.69$ to $83.61$ps. Using the Vernier effect, we further reduced the resolution. Figure~\hyperref[fig6]{6e-f} illustrate the temporal responses of the coupled cavities with different lengths, achieving sampling times as low as $7.52$ to $20.9$~ps. This reduction is especially beneficial for applications requiring extremely high delay precision.

\begin{table}[h]
\caption{Parameters of the Vernier cavities}\label{tab1}
\begin{tabular*}{\textwidth}{@{\extracolsep\fill}lccccccccc}
\toprule%
& \multicolumn{3}{@{}c@{}}{FSR$\times 3.6$} 
& \multicolumn{3}{@{}c@{}}{FSR$\times 6$} 
& \multicolumn{3}{@{}c@{}}{FSR$\times 10$} 
\\\cmidrule{2-4}\cmidrule{5-7}\cmidrule{8-10}%

Parameter & R1 & R2 & Total & R1 & R2 & Total & R1 & R2 & Total \\
\midrule
L (mm)      & 6 & 4.5 &  & 5.4 & 4.5 &  & 5.4 & 4.86 &  \\
$\alpha$    & 0.8639 & 0.9112 &  & 0.8478 & 0.9112 &  & 0.8478 & 0.8892 &  \\
\midrule
$\kappa_1,\kappa_3$     &  &  & 0.3285 &  &  & 0.3714 &  &  & 0.2282 \\
$\kappa_2$              &  &  & 0.1122 &  &  & 0.1530 &  &  & 0.1428 \\
$\text{P}_\text{T}$\footnotemark[1](mW) &  &  & 7.45 &  &  & 8.92 &  &  & 10.15 \\
\midrule
FSR(GHz)   & 11.96 & 15.95 & 47.84 & 13.29 & 15.95 & 79.74 & 13.29 & 14.76 & 132.9 \\
T (ps)      & 83.61 & 62.69 & 20.90 & 75.24 & 62.69 & 12.54 & 75.24 & 67.75 & 7.52 \\
$\varphi_\text{CT}$ (rad) & $0.28\pi$ & 0 & $0.67\pi$ & $0.28\pi$ & 0 & $0.4\pi$ & $0.28\pi$ & 0 & $0.22\pi$ \\
$\varphi_\text{FT}$ (rad) & $0.28\pi$ & $0.28\pi$ & $0.133\pi$ & $0.28\pi$ & $0.28\pi$ & $0.08\pi$ & $0.28\pi$ & $0.28\pi$ & $0.002\pi$ \\
\botrule
\end{tabular*}
\footnotetext{The total phase shift $\varphi$ is calculated on the total FSR resulting from the Vernier effect. CT: coarse tuning, FT: fine tuning.}
\footnotetext[1]{Total power consumption of the ciruit.}
\end{table}

\section{Discussion}
In this work, we have proposed, fabricated and experimentally demonstrated a non-uniform programmable photonic mesh in silicon on insulator. Specifically, we introduced defects within a hexagonal mesh structure to break the uniformity of the cavities and path length sections. By coupling two or more cavities with different round-trip lengths, we successfully implemented filters with an extended FSR, leveraging the Vernier effect to enhance performance. This approach addresses a critical limitation in uniform programmable hexagonal meshes, where each cavity comprises six MZIs, leading to long ring resonators with inherently small FSRs. 

To validate our concept, we designed and fabricated a compact proof-of-concept device featuring four hexagonal cavities, two of which incorporate an internal MZI that splits the cavity. Using this programmable mesh, we demonstrated three different coupled-cavity filters with extended FSRs, achieving ring resonator responses with FSR values of $47.8$, $79.7$, and $132.9$~GHz. In contrast, a uniform mesh of the same fabricated size yields a maximum FSR of only $13.3$~GHz, indicating a tenfold improvement. Furthermore, our approach enables a significantly reduced time sampling rate, which is crucial for high-speed optical signal processing. We obtained time sampling values of $20.9$, $12.5$, and $7.5$~ps using the Vernier cavities, compared to approximately $75$~ps when utilizing a single hexagonal cavity.

This study paves the way for further advancements in non-uniform programmable photonic meshes. A promising direction is scalability, where larger meshes incorporating multiple defect cells could be designed to achieve even higher FSR multiplication factors, reducing the size of the TBU and improving overall system efficiency. The current TBU footprint of $900$~$\upmu$m could be reduced to approximately $100$~$\upmu$m, enabling operation bandwidths of around $120$~GHz for hexagonal cavities and up to $1.2$~THz for the proposed non-uniform mesh \cite{Perez2024,Yingjie2019}. Additionally, minimizing TBU losses will further enhance the performance of the Vernier effect, leading to higher extinction ratios in the main resonances. The current loss of approximately $0.48$~dB per TBU could potentially be reduced to around $0.08$~dB \cite{lightelligence}, enabling applications in ultra-broadband optical signal processing and high-speed optical communications. 

Beyond classical applications, non-uniform photonic meshes provide a highly adaptable platform for emerging fields such as topological photonic devices \cite{Dai2024} and quantum photonics \cite{Dong2023}, offering new opportunities for advanced photonic computing and signal processing architectures.

\section{Methods}
\subsection{Fabrication of the device}
The photonic integrated circuit was fabricated by Advanced Micro Foundry using a standard silicon on insulator process. The chip was produced from an SOI wafer with a 220 nm device layer, featuring 500 nm single-mode waveguides patterned through deep ultraviolet (193 nm) lithography. Phase-shifting waveguide sections were implemented by depositing a 120 nm TiN heater layer over the waveguides, powered by 2000 nm-thick metal DC tracks added in the final stages of fabrication. The process allows the doping of waveguides and under-etched waveguides to improve the efficiency of the phase actuators and mitigate thermal crosstalk effects. The chip size is $2.8 \times 4.3$~mm$^2$ including electrical pads and edge couplers.

\subsection{Device characterization and measurements}
Manufacturing deviations affect the optical phase of the elements that compound the programmable mesh. Characterization of all the MZIs is needed to correctly drive the currents at each thermal tuner. This process was performed using a continuous wave laser fixed at $1550$~nm and a power meter (FTBx-1750 EXFO) to measure the output optical power while driving the phase shifters. The 42~DC channels are wirebonded to a printed circuit board (PCB) and controlled with a multichannel current source (Qontrol Ltd.) that can be programmed through coding. A more detailed explanation of the calibration process can be found in Supplementary Note 4. After calibration, each TBU has its own set of current values for a certain coupling and phase.

For optical spectral measurements, we used a tunable laser source (T100S-HP EXFO) along with a component tester (CT440 EXFO), featuring a 1~pm wavelength resolution. Active optical alignment was performed automatically using piezo controllers (Thorlabs MDT693B) for the positioner that held the fibre array. The alignment routine consists of finding, simultaneously, high optical output power at both loops of the edge coupler array. A 7~dB loss is measured for the loop ports, corresponding to a 3.5 dB loss per input, including the coupling and insertion losses of the edge couplers. 

\backmatter

\bmhead{Supplementary information}
See the attachment file.

\bmhead{Acknowledgements} The authors acknowledge the financial support of the European Research Council through the project ERC-POC-2023-101138302 NU-MESH and the Advanced
Grant program under grant Agreement No. 101097092 (ANBIT). Also, the COMCUANTICA/005
and COMCUANTICA/006 grants, funded by the European Union through NextGenerationEU (PRTR-C17) with the support of the Spanish Ministry of Science and Innovation and the Generalitat Valenciana. Finally, the project PROMETEO/2021/015 funded by the Generalitat Valenciana.

\bmhead{Author Contributions} J.C. conceived the idea of the project. J.C. and C.C. carried out the design of the non-uniform mesh. C.C. simulated and designed the silicon chip, and performed the calibration and measurements of the fabricated mesh. J.C. and C.C. wrote the paper. J.C. supervised the project.

\section*{Declarations}

The authors declare no conflicts of interest.

\bibliography{sn-bibliography}

\newpage
\setcounter{page}{1}
\begin{center}
    \LARGE{Non-uniform programmable photonic waveguide meshes} 
    \\ \vspace{0.5 cm}
    \Large{Supplementary Notes}
\end{center}

\section*{Contents}
\textbf{Supplementary Note 1.} Modelling and optimization of the Vernier effect \dotfill \pageref{sec1} \\
\textbf{Supplementary Note 2.} Introducing defect cells in hexagonal meshes\dotfill \pageref{sec2} \\
\textbf{Supplementary Note 3.} Flattened hexagonal mesh design\dotfill \pageref{sec3} \\
\textbf{Supplementary Note 4.} Characterization and calibration of the device\dotfill \pageref{sec4} \\
\textbf{Supplementary Note 5.} Setup preparation and measurement\dotfill \pageref{sec5} \\
\textbf{Supplementary Note 6.} Characterization of the different cells\dotfill \pageref{sec6} \\
\textbf{Supplementary Note 7.} Kramers-Kronig relations for phase response\dotfill \pageref{sec7} \\
\textbf{Supplementary References} \dotfill \pageref{ref} \\

\newpage
\section*{Supplementary Note 1. \normalfont Modelling and optimization of the Vernier effect}\label{sec1}

A serially coupled double-ring resonator enables the extension of the free spectral range (FSR) to the least common multiple of the FSR of the individual rings \cite{ref1}. This is achieved by designing the two rings with different radii. When light propagates through the double-ring resonator, it is transmitted from the drop port only when both rings meet their respective resonance conditions. The resulting FSR of the double-ring resonator with distinct radii can be expressed as:
\begin{equation}
    \text{FSR}_\text{T} = n \text{FSR}_1 = m \text{FSR}_2
\end{equation}
Multiplying both sides of the first by $\text{FSR}_2$ and both sides of the second by $\text{FSR}_1$, we get: 
\begin{equation}
    \text{FSR}_\text{T} \text{FSR}_2 = n \text{FSR}_1 \text{FSR}_2
\end{equation}
\begin{equation}
    \text{FSR}_\text{T} \text{FSR}_1 = m \text{FSR}_2 \text{FSR}_1
\end{equation}
Then, we subtract both equations to get:
\begin{equation}
    \text{FSR}_\text{T} \cdot |\text{FSR}_1 - \text{FSR}_2| = |m-n| \cdot \text{FSR}_1  \text{FSR}_2
\end{equation}
Which leads to:
\begin{equation}
    \text{FSR}_\text{T} = |m-n|\dfrac{\text{FSR}_1 \cdot \text{FSR}_2}{|\text{FSR}_1 - \text{FSR}_2|}
\end{equation}
Where $n$ and $m$ are natural and coprime numbers, so $|m-n| = 1$:
\begin{equation}
    \text{FSR}_\text{T} = \dfrac{\text{FSR}_1 \cdot \text{FSR}_2}{|\text{FSR}_1 - \text{FSR}_2|}
\end{equation}

\begin{figure}[h]
\centering
\includegraphics[width=0.4\textwidth]{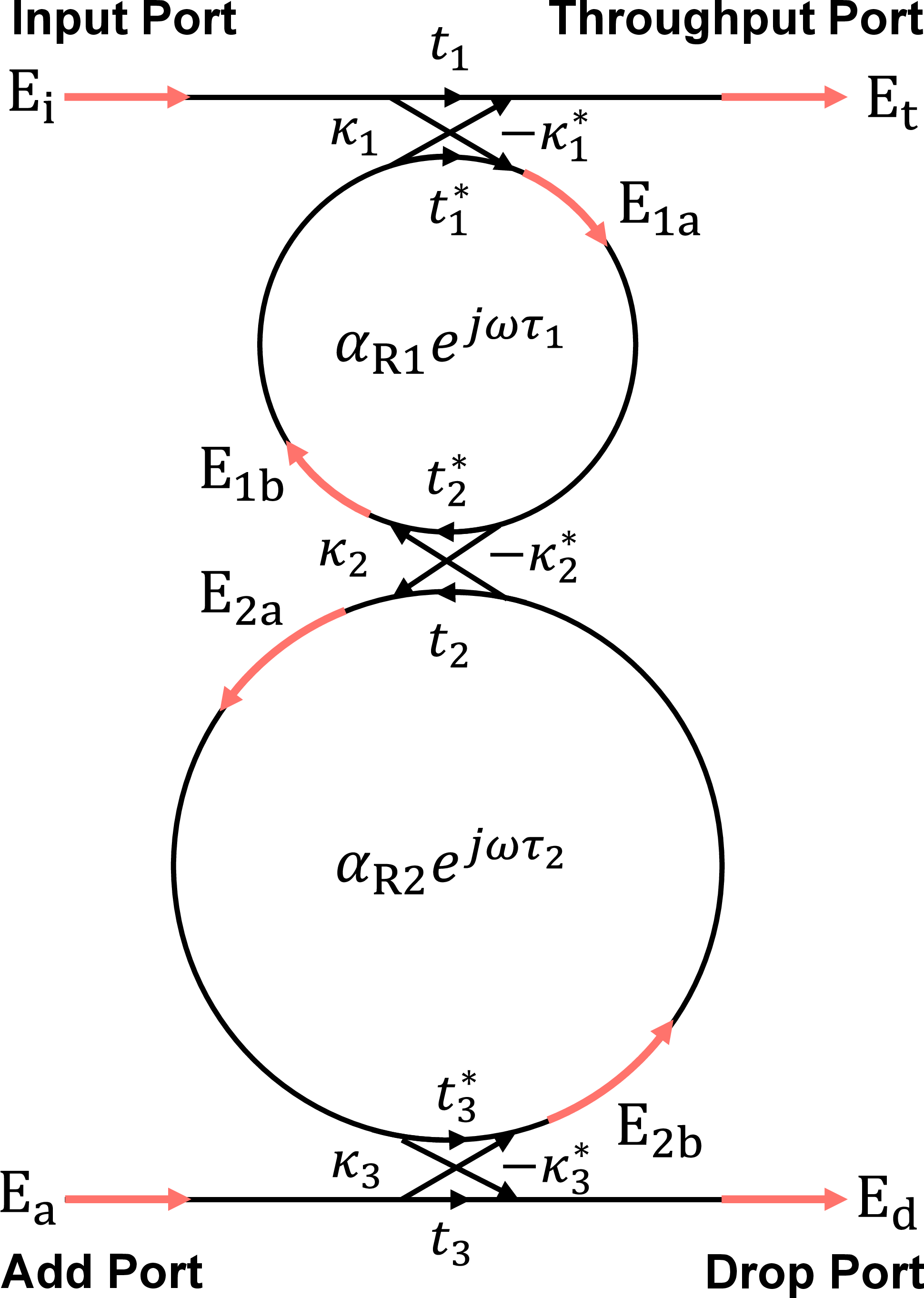}
\caption{Serially coupled double-ring resonator}
\label{fig1s}
\end{figure}

The schematic of a serially coupled double-ring resonator is depicted in Fig.~\ref{fig1s}.
From this model, the fields can be calculated as follows:
\begin{equation}
    E_{1a} = -\kappa_1^* E_{i} + t_1^* \alpha_1 e^{j\omega\frac{\tau_1}{2}} E_{1b}
\end{equation}
\begin{equation}
    E_{1b} = t_2^* \alpha_1 e^{j\omega\frac{\tau_1}{2}} E_{1a} -\kappa_2^* \alpha_2 e^{j\omega\frac{\tau_2}{2}} E_{2b}
\end{equation}
\begin{equation}
    E_{2a} = \kappa_2 \alpha_1 e^{j\omega\frac{\tau_1}{2}} E_{1a} + t_2 \alpha_2 e^{j\omega\frac{\tau_2}{2}} E_{2b}
\end{equation}
\begin{equation}
    E_{2b} = -\kappa_3^* E_{a} + t_3^* \alpha_2 e^{j\omega\frac{\tau_2}{2}} E_{2a}
\end{equation}
\begin{equation}
    E_t = t_1 E_{i} + \kappa_1 \alpha_1 e^{j\omega\frac{\tau_1}{2}} E_{1b}
\end{equation}
\begin{equation}
    E_d = t_3 E_{a} + \kappa_3 \alpha_2 e^{j\omega\frac{\tau_2}{2}} E_{2a}
\end{equation}
Where $\alpha_1 = \alpha_{\text{R}1/2}$ and $\alpha_2 = \alpha_{\text{R}2/2}$ represent the loss coefficients of the half round-trip length of the rings, $\omega = 2\pi f$ is the frequency vector, and $\tau = L/c$ is the round trip delay. From (7) to (12), the general expressions for the transfer functions for the throughput and the drop port can be derived. Here we assume a coupler without losses and symmetric coupling behaviour, so $t=t^*$ and $\kappa = \kappa^*$ (note that $|t^2| + |\kappa^2| = 1$). Assuming that there is no input field in the add port ($E_a = 0$), we obtain: 

\begin{equation}
    \dfrac{E_t}{E_i} = \dfrac{-t_1 \kappa_1^2 \alpha_1^2  e^{j\omega \tau_1} (t_3 \alpha_2^2 e^{j\omega \tau_2} - t_2)}
    {1 - t_1 t_2 \alpha_1^2 e^{j\omega \tau_1} - t_2 t_3 \alpha_2^2 e^{j\omega \tau_2}  + t_1 t_3 \alpha_1^2 \alpha_2^2 e^{j\omega \tau_1} e^{j\omega \tau_2}  }
\end{equation}

\begin{equation}
    \dfrac{E_d}{E_i} = \dfrac{\kappa_1 \kappa_2 \kappa_3 \alpha_1 \alpha_1 e^{j\frac{\omega \tau_1}{2}} e^{j\frac{\omega \tau_2}{2}}}{1 - t_1 t_2 \alpha_1^2 e^{j\omega \tau_1} - t_2 t_3 \alpha_2^2 e^{j\omega \tau_2}  + t_1 t_3 \alpha_1^2 \alpha_2^2 e^{j\omega \tau_1} e^{j\omega \tau_2}  }
\end{equation}
\vspace{0.5 cm}

If we have a serially coupled structure of $N$ rings, where each ring has a different length and therefore a different FSR, we could deduce its total FSR recursively:

\begin{equation}
    \text{FSR}_{\text{T}_1} = \dfrac{\text{FSR}_1 \cdot \text{FSR}_2}{|\text{FSR}_1 - \text{FSR}_2|}
\end{equation}

\begin{equation}
    \text{FSR}_{\text{T}_2} = \dfrac{\text{FSR}_{\text{T}_1} \cdot \text{FSR}_3}{|\text{FSR}_{\text{T}_1} - \text{FSR}_3|} = 
    \dfrac{\text{FSR}_1 \cdot \text{FSR}_2 \cdot \text{FSR}_3}{|\text{FSR}_1 - \text{FSR}_2| \cdot \left| \dfrac{\text{FSR}_1 \cdot \text{FSR}_2}{|\text{FSR}_1 - \text{FSR}_2|} -  \text{FSR}_3\right|}
\end{equation}

This brings us to the following equation for structures with $N>2$:

\begin{equation}
    \text{FSR}_{\text{T}_N} = \text{FSR}_{\text{T}_1} \cdot \prod_{i=3}^{N} \dfrac{\text{FSR}_i}{\left| \text{FSR}_i - \text{FSR}_{\text{T}_{i-2}} \right|}
\end{equation}

\newpage
\section*{Supplementary Note 2. \normalfont Introducing defect cells in hexagonal meshes}\label{sec2}
In a programmable photonic mesh, it is possible to introduce defects or non-uniformities in different ways, allowing for controlled modifications to the structure and functionality of the system \cite{ref2}. Below, we will describe two specific cases where the overall hexagonal perimeter of the mesh is preserved. In the first case, we introduce one or multiple Mach-Zehnder Interferometers (MZIs) inside the hexagons. In the second case, we shorten one of the hexagonal sides.

\begin{figure}[h]
\centering
\includegraphics[width=\textwidth]{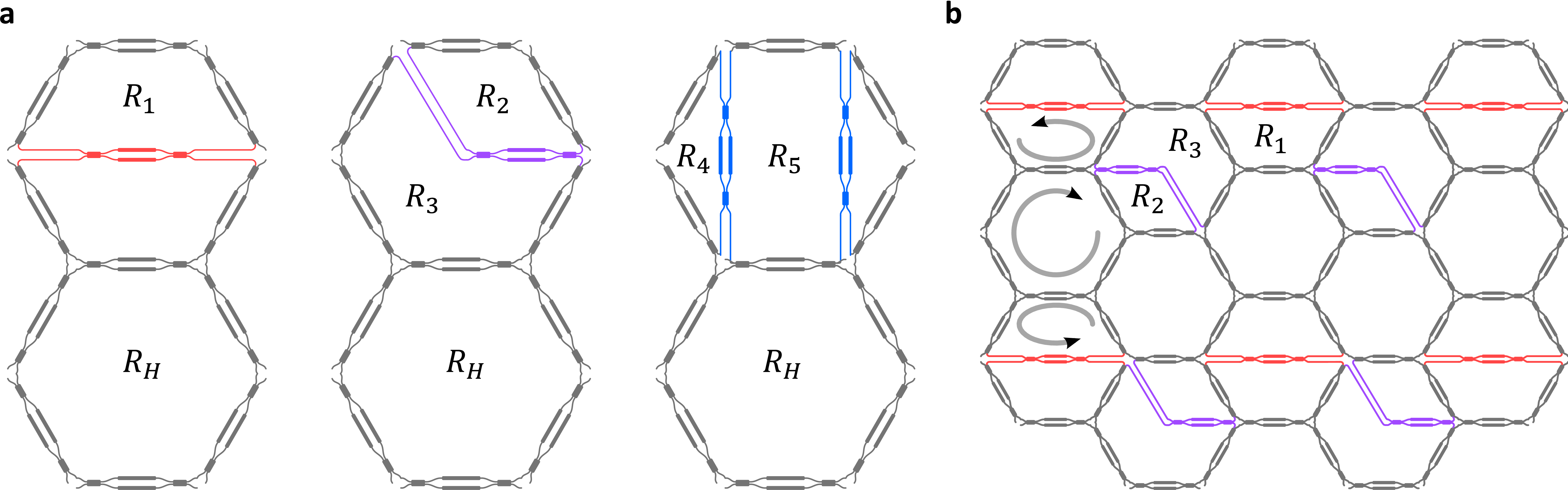}
\caption{(a) Different configurations of defect cells introduced into a hexagonal mesh. The defects modify the internal structure while maintaining the external hexagonal perimeter. (b) Example of a scalable non-uniform programmable mesh with strategically placed defects.}
\label{fig2s}
\end{figure}

The first case is the one discussed in this work, where a defect MZI is introduced inside the hexagonal cell. Figure~\ref{fig2s}a shows an example of possible configurations (note that more combinations are possible). The first defect, shown in red in the leftmost panel, represents a horizontal modification within the mesh. The second defect, highlighted in purple in the second panel, introduces modification with two different cavities inside the hexagonal cell. Finally, the third defect depicted in blue, forms a vertical variation with three cavities inside the cell, using two MZIs.

These defects can be distributed within the programmable hexagonal framework. Figure~\ref{fig2s}b shows an example of how we can scale a programmable mesh with defects. The modifications are applied in alternating rows to create a structured non-uniformity, where all the cavities can be connected with the hexagonal one.

\begin{figure}[h]
\centering
\includegraphics[width=\textwidth]{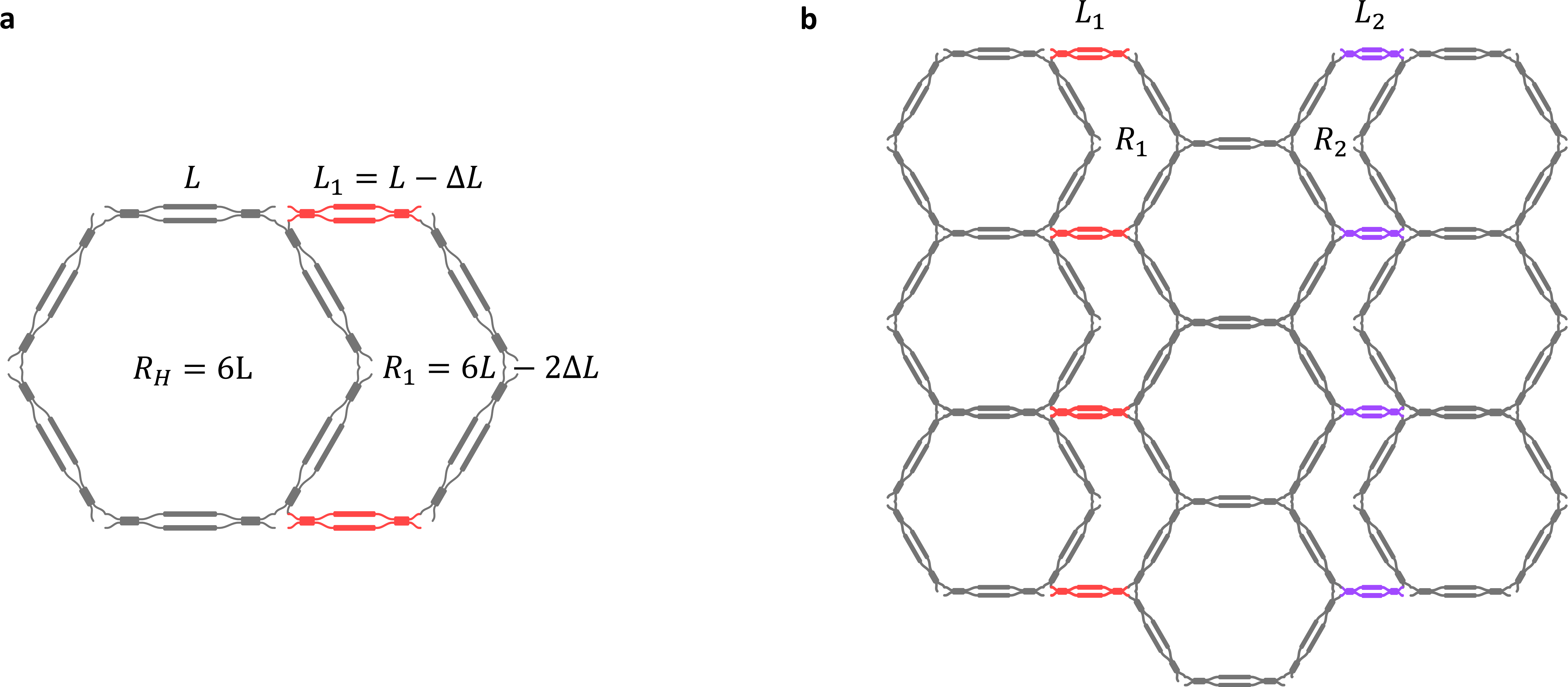}
\caption{(a) A hexagonal cell with a defect cell placed in series. (b) Large-scale implementation of the defect-cell approach, showing periodic non-uniformities across the structure in columns.}
\label{fig3s}
\end{figure}

Figure~\ref{fig3s}a represents another configuration of a non-uniform programmable photonic mesh, where specific modifications are introduced to alter the structure and optical behaviour of the hexagonal network. A single hexagonal unit is shown with local changes. The red-highlighted elements indicate the introduction of Mach-Zehnder Interferometers (MZIs) along the side edges of the hexagon, with a shorter length, forming an extension of the hexagon itself.

Figure~\ref{fig3s}b shows a larger section of the photonic mesh, where these modifications are systematically arranged. The defect cells are distributed in columns, creating a periodic pattern that introduces non-uniformities in the structure. 

\begin{figure}[h]
\centering
\includegraphics[width=\textwidth]{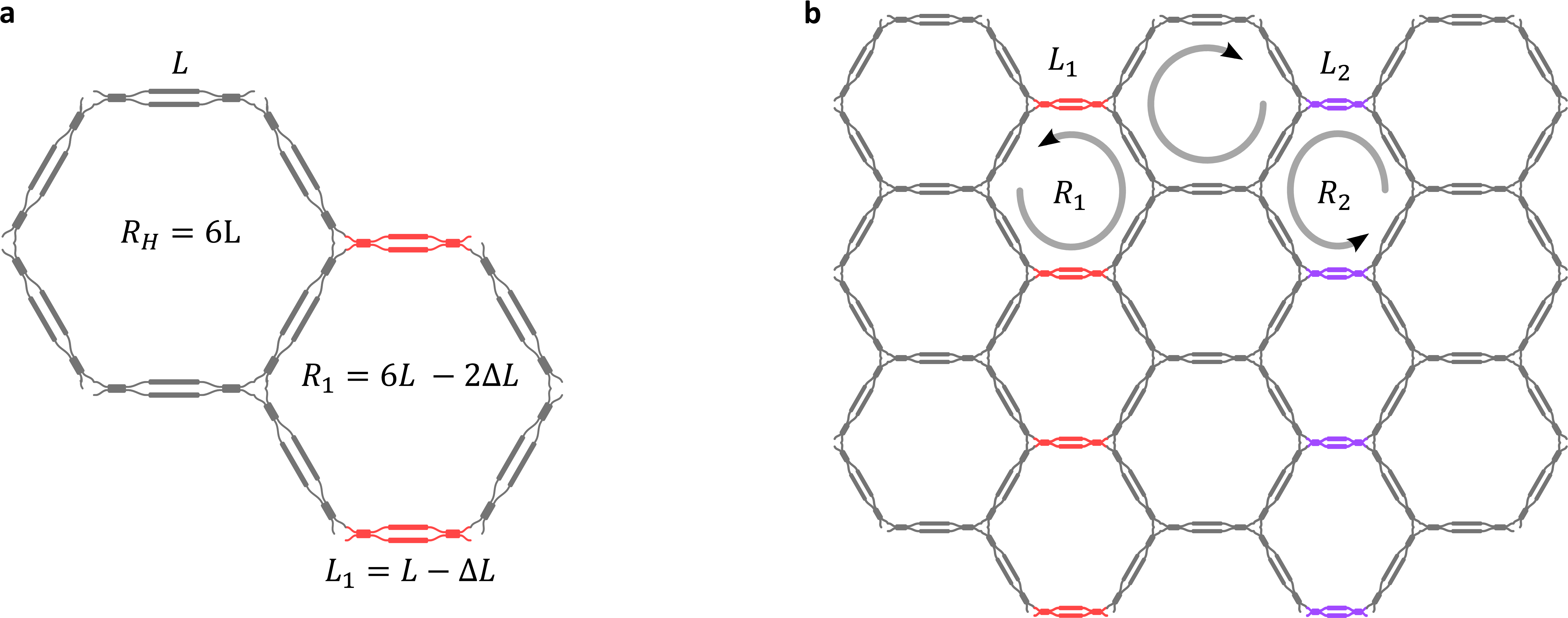}
\caption{(a) Alternative defect-cell configuration preserving original uniform mesh connections. (b) Column-based distribution of defect cells to maintain connectivity while introducing structured non-uniformities.}
\label{fig4s}
\end{figure}

One of the problems with adding the defect as shown in Fig.~\ref{fig3s} is that we waste one of the connections of the hexagons adjacent to the defect cells. Therefore, one way to solve this would be to follow this structure but keep the connections of the original uniform mesh (Fig.~\ref{fig4s}a). In this case, we must also introduce the defect cells in columns as shown in Fig.~\ref{fig4s}b. 

\newpage
\section*{Supplementary Note 3. \normalfont Flattened hexagonal mesh design}\label{sec3}
\begin{figure}[h]
\centering
\includegraphics[width=0.9\textwidth]{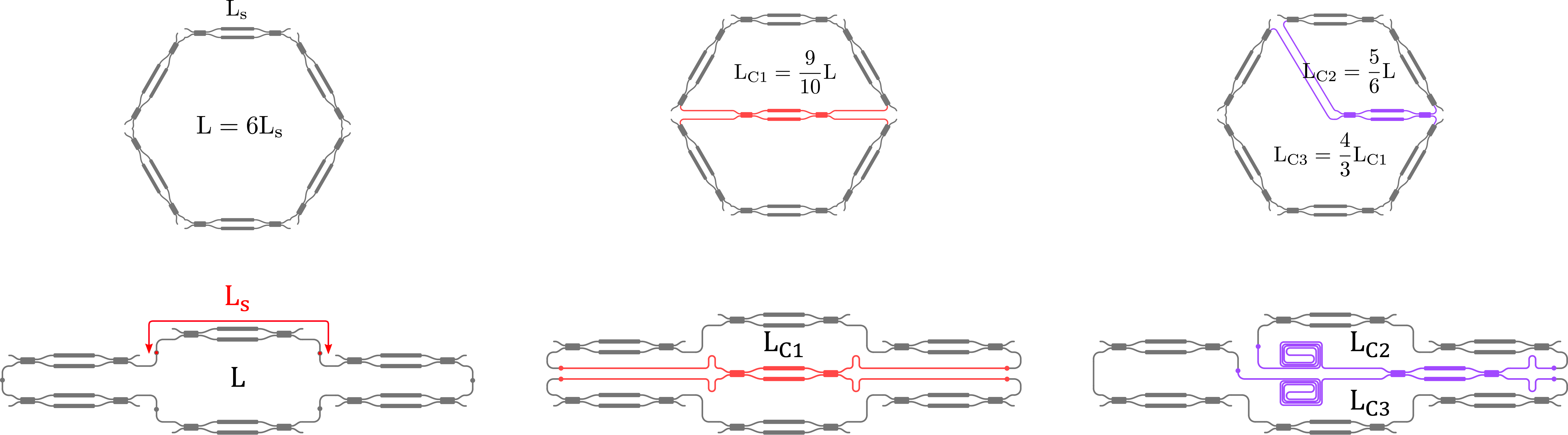}
\caption{Flattened hexagonal mesh designs optimized for space efficiency. The modified layout aligns all MZIs at the same angle, minimizing footprint while maintaining functionality.}
\label{fig5s}
\end{figure}

For the design of the layout of the fabricated chip, the available space was limited and therefore, the hexagonal cells have been modified so that they take up less space on one of the axes. To do this, the hexagonal shape has been "flattened", making all the MZI positioned at the same angle \cite{ref3}. Figure~\ref{fig5s} shows the three cell designs presented in this work in their flatten-like form.

\begin{figure}[h]
\centering
\includegraphics[width=0.9\textwidth]{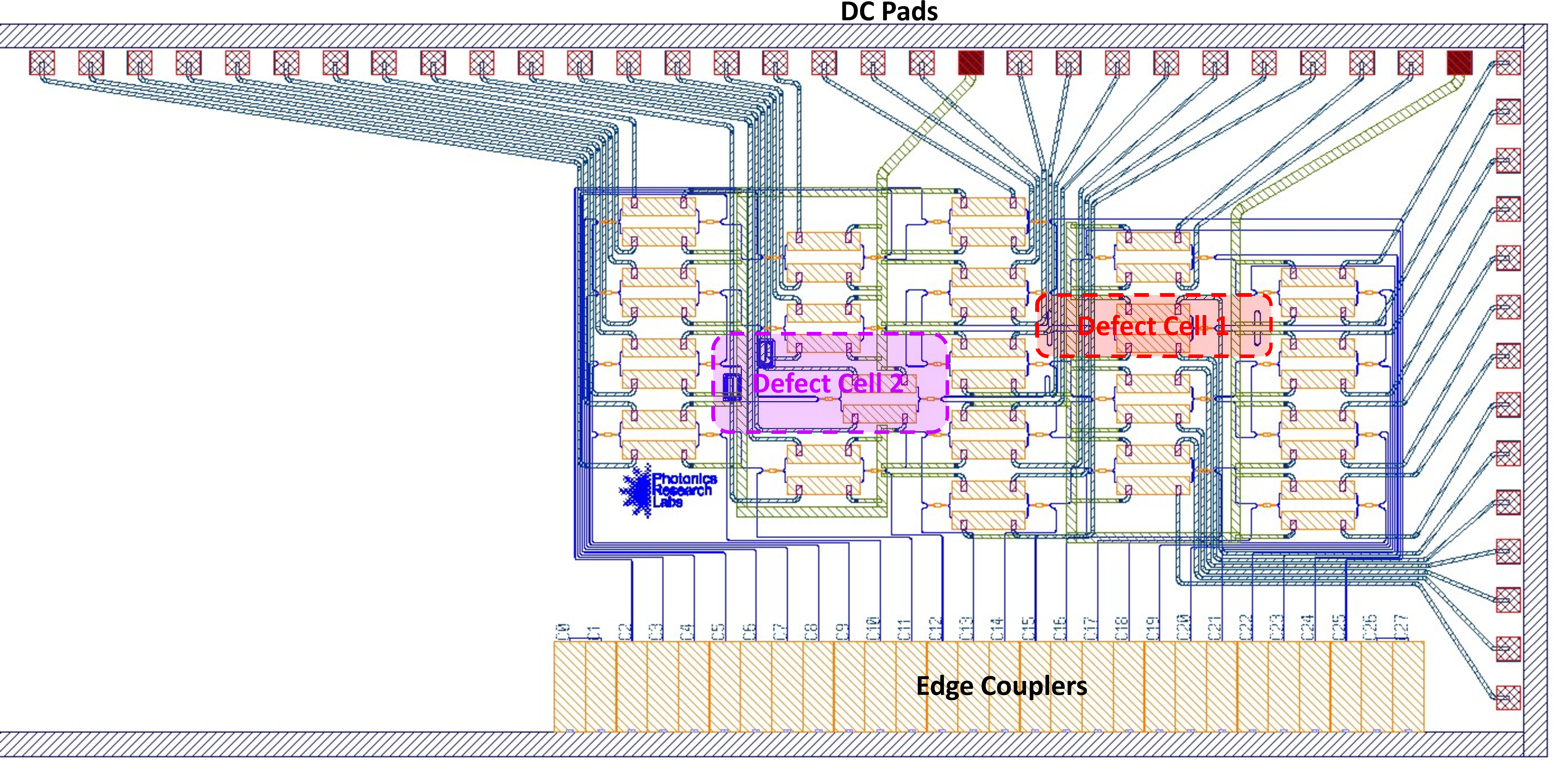}
\caption{Complete layout of the non-uniform programmable photonic mesh, including edge couplers, DC pads, and metal tracks. MZIs in the defect cells are highlighted in red and purple.}
\label{fig6s}
\end{figure}

In this type of configuration, it is necessary to add an extra length to each side of the MZI to make the joints between them. This makes the length of each side of the hexagon Ls, inevitably a little longer than in the traditional hexagonal design. In the case of defects, it has been necessary to adjust the lengths by adding extra paths or spirals, to obtain the required roundtrip of the ring to create the Vernier effect.
The complete layout of the non-uniform mesh is shown in Fig.~\ref{fig6s}, including edge couplers, DC pads and metal tracks. The MZIs in the default cells are highlighted in red and purple.

\newpage

\section*{Supplementary Note 4. \normalfont Characterization and calibration of the device}\label{sec4}
\begin{figure}[h]
\centering
\includegraphics[width=\textwidth]{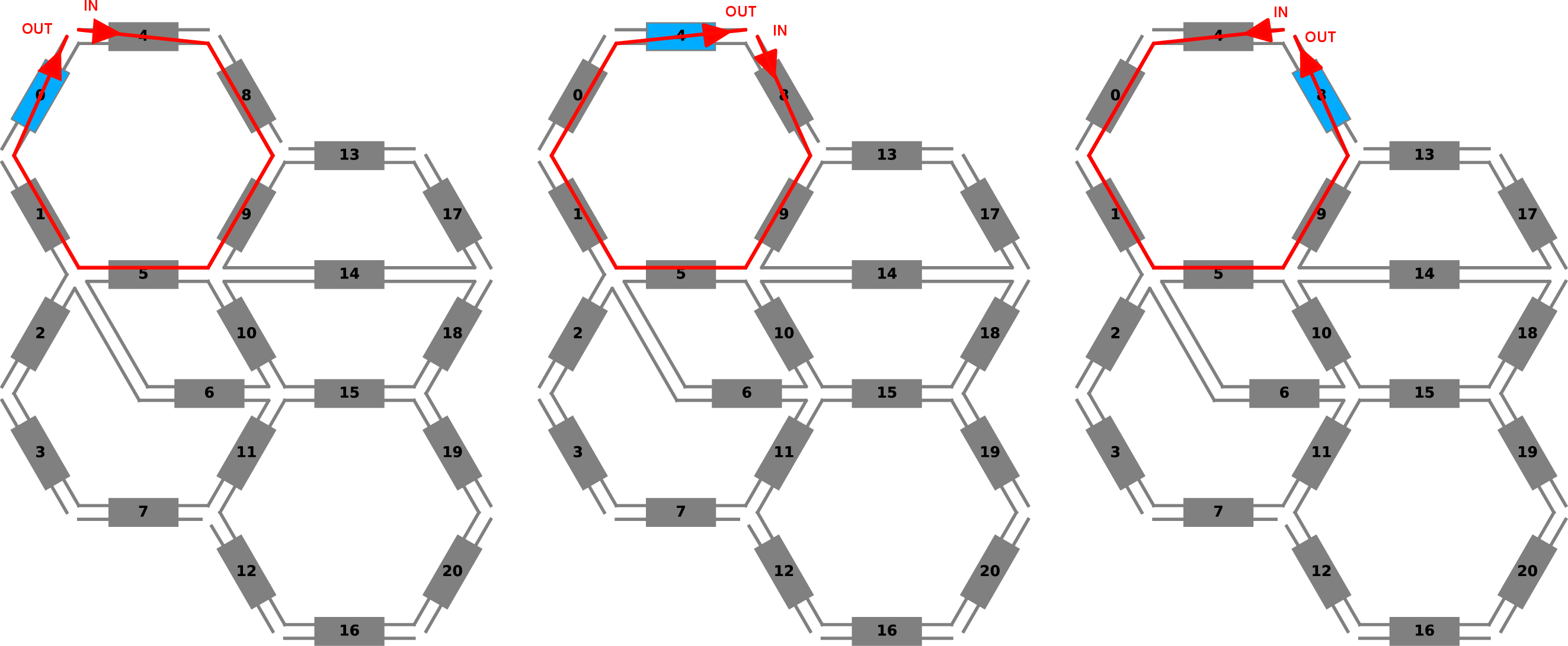}
\caption{Examples of configurations for MZI characterization in the perimeter of the hexagonal mesh. The optical paths used for characterization are highlighted in red.}
\label{fig7s}
\end{figure}

The characterization of the MZI of the mesh was carried out manually since in this case, it is a small mesh. Other calibrations based on graphs can be used for automatic calibration \cite{ref4, ref5}. In this case, we have used a fixed laser at $1550$~nm and a power meter to measure the optical power at the output.
First, a characterization of the MZIs that are in the perimeter of the mesh has been carried out, since having only 2 of 4 ports connected, their characterization is simpler. To do this, we have followed a recirculating characterization scheme. Fig.~\ref{fig7s} shows 3 examples of MZI that are located on the perimeter of the hexagonal cell on the left, with the MZI to be characterized in blue and the optical path followed in red. 
By recirculating scheme, we imply that we use as an output port the MZI that we want to characterize and any input port that has direct access to the cell where that MZI is located. Assuming that enough light reaches the MZI that we are characterizing, we sweep each of its phase shifters in current and measure the optical power at the output. With this, we will have characterized the MZI without any interfering signal on the other input port.

Once the MZIs from the perimeter of the mesh have been characterized, we can characterize the inner MZIs by following the same technique and putting the adjacent MZI in a \textit{cross} state so that it interferes.

\begin{figure}[h]
\centering
\includegraphics[width=0.8\textwidth]{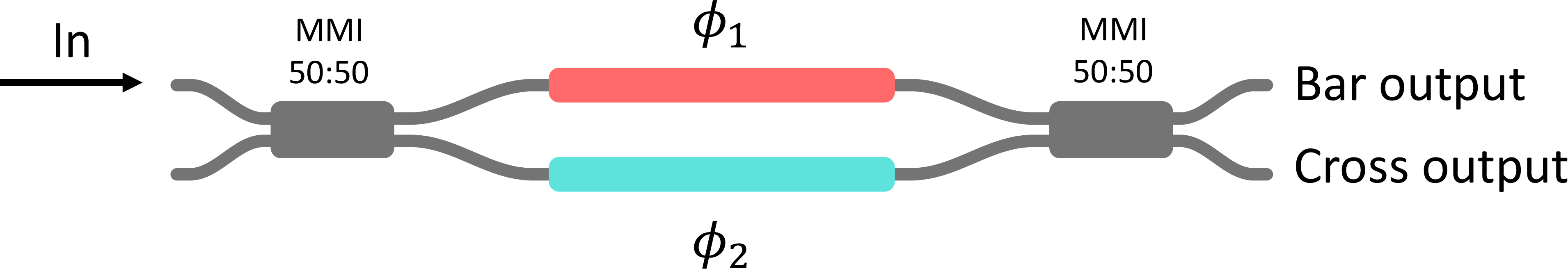}
\caption{Schematic representation of an MZI implementation using two 3-dB couplers, with tunable phase shifters for independent control of phase and coupling.}
\label{fig8s}
\end{figure}

The MZI are implemented by using two 3-dB couplers, where each arm incorporates a tunable phase shifter as shown in Fig.~\ref{fig8s}. 
Considering MMIs for the 3-dB couplers \cite{ref6}, the transmission matrix results in:
\begin{equation}
\begin{split}
    T &= \dfrac{je^{j\varphi_0}}{\sqrt{2}}\dfrac{je^{j\varphi_0}}{\sqrt{2}} 
    \begin{pmatrix}
        1 & j \\
        j & 1
    \end{pmatrix}
    \begin{pmatrix}
        e^{-j\phi_1} & 0 \\
        0 & e^{-j\phi_2}
    \end{pmatrix}
    \begin{pmatrix}
        1 & j \\
        j & 1
    \end{pmatrix}
    = \\
    &= \dfrac{e^{j(2\varphi_0 + \pi)}}{2} 
    \begin{pmatrix}
        1 & j \\
        j & 1
    \end{pmatrix}
    \begin{pmatrix}
        e^{-j\phi_1} & je^{-j\phi_1} \\
        je^{-j\phi_2} & e^{-j\phi_2}
    \end{pmatrix}
    = \\
    &= \dfrac{e^{j(2\varphi_0 + \pi)}}{2} 
    \begin{pmatrix}
        e^{-j\phi_1} - e^{-j\phi_2} & je^{-j\phi_1} + je^{-j\phi_2} \\
        je^{-j\phi_1} + je^{-j\phi_2} & -e^{-j\phi_1} + e^{-j\phi_2}
    \end{pmatrix}
    = \\
    &= e^{j\left(2\varphi_0 - \frac{\phi_1 + \phi_2 - 3\pi}{2}\right)}
    \begin{pmatrix}
        \sin \left( \frac{\phi_2 - \phi_1}{2}\right) & \cos \left( \frac{\phi_2 - \phi_1}{2}\right) \\
        \cos \left( \frac{\phi_2 - \phi_1}{2}\right) & -\sin \left( \frac{\phi_2 - \phi_1}{2}\right)
    \end{pmatrix},
\end{split}
\end{equation}

\noindent where $\varphi_0 = -\beta_0 L_{\text{MMI}} - 3\pi/4$ is the phase corresponding to the $2\times2$ MMIs.

We can define the final transfer matrix, considering the overall phase and coupling as follows:
\begin{equation}
    \varphi = -\dfrac{\phi_1 + \phi_2}{2}
\end{equation}
\begin{equation}
    \theta = \dfrac{\phi_2 - \phi_1}{2}
\end{equation}
Which leads to: 
\begin{equation}
    T = -je^{j\varphi}
    \begin{pmatrix}
        \sin \theta & \cos \theta \\
        \cos \theta & -\sin \theta
    \end{pmatrix},
\end{equation}
Equation (18) represents a tunable coupler where the coupling constant $\kappa = \cos^2 \theta$ and the overall phase shift $\varphi$ can be independently tuned according to the system of two equations given by (16) and (17).

\begin{figure}[h]
\centering
\includegraphics[width=0.65\textwidth]{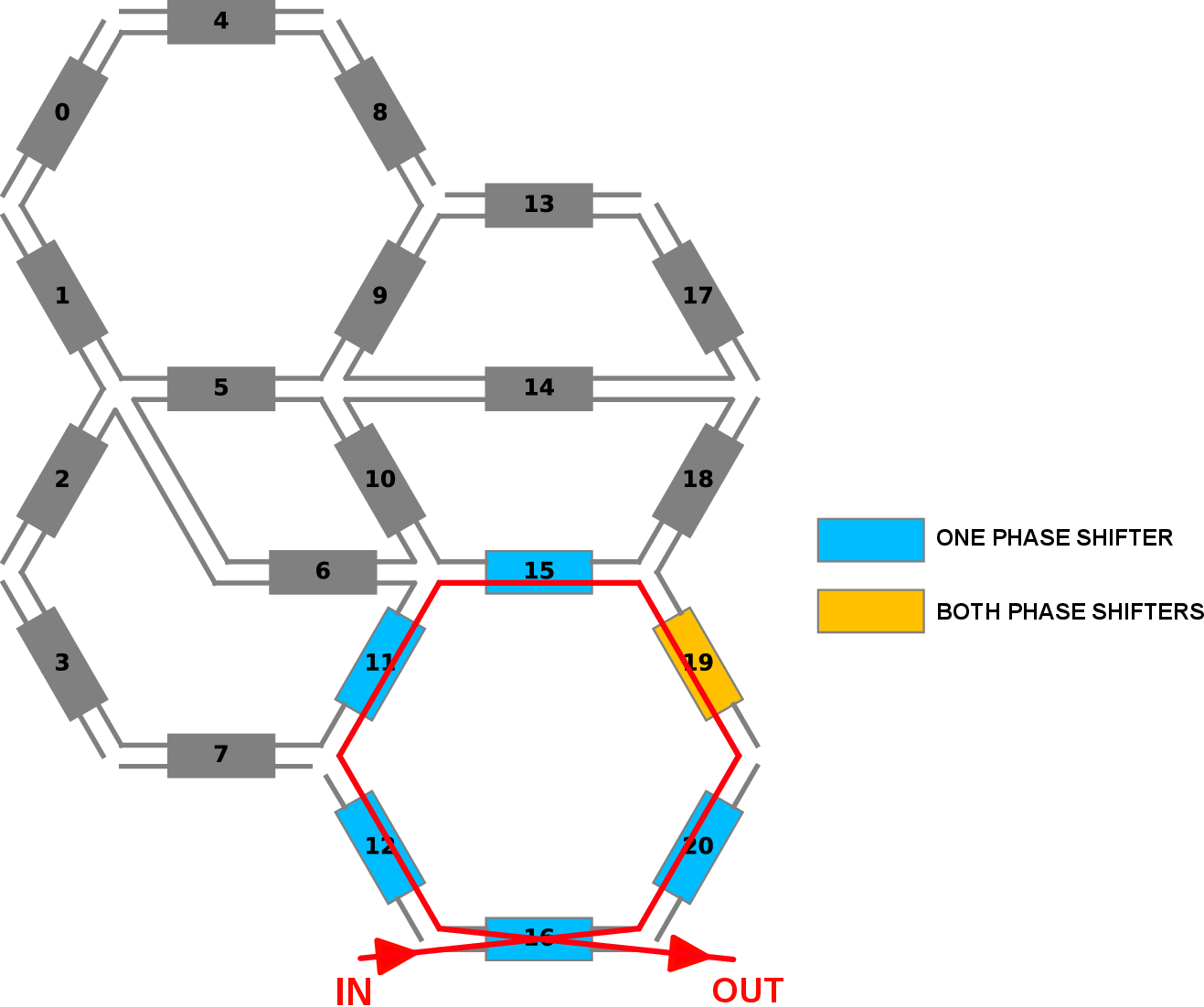}
\caption{Example of a tunable ring resonator implemented within the programmable mesh. Specific MZIs are controlled to define the coupling factor and phase tuning.}
\label{fig9s}
\end{figure}

\begin{figure}[h]
\centering
\includegraphics[width=0.8\textwidth]{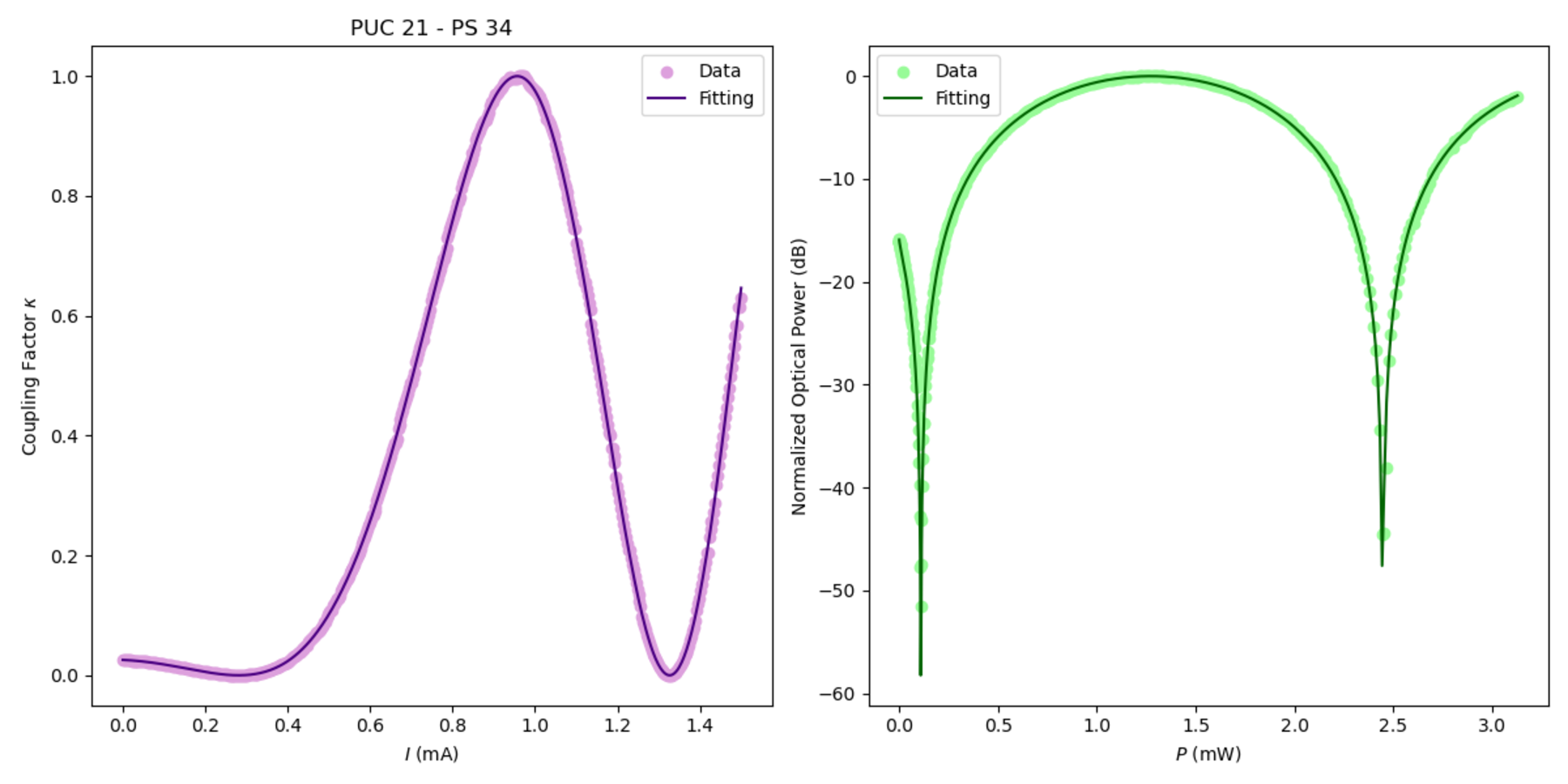}
\caption{Measured coupling as a function of applied current (left) and transmission response in dB (right). The experimental data is fitted using a polynomial model to account for fabrication deviations.}
\label{fig10s}
\end{figure}

For example, for implementing a tunable ring resonator in the mesh, an example in Fig.\ref{fig9s} is shown. Here we use phase shifter 16 for coupling the light to the ring with a coupling factor $\kappa$ driving only one of the phase shifters of that MZI. Currents of the MZIs 11, 12, 15, 19 and 20 are set to provide a coupling factor $\kappa = 1.0$ (i.e., bar state). Blue MZIs are also controlled using only one phase shifter. Then, MZI number 19 (depicted in orange) is set using both phase shifters. In that way, we can maintain the bar state and also tune the phase of the ring. To do that, we set the current $i_0$ of one of the phase shifters ($\phi_1$ for example) for having the coupling $\kappa = 1.0$ and then we can tune the phase of the ring changing the current of the two phase shifters as following:
\begin{equation}
    i_1 = \sqrt{i_0^2 + i_r^2}
\end{equation}
\begin{equation}
    i_2 = i_r,
\end{equation}
where $i_1$ and $i_2$ are the current of each phase shifter and $i_r$ the additional current to control the phase.
Figure~\ref{fig10s} shows the measurement of a characterized MZI, where coupling can be observed as a function of current (left side) and also the measurement in dB as a function of the power applied to the phase shifter (right side). In addition, the fitting performed using the above equations can be observed, where $\theta = p_0 + p_1I + p_2 I^2 + p_3 I^3 + p_4 I^4$. Higher orders than $I^2$ were used to take into consideration fabrication deviations.

\newpage
\section*{Supplementary Note 5. \normalfont Setup preparation and measurement}\label{sec5}
\begin{figure}[h]
\centering
\includegraphics[width=0.9\textwidth]{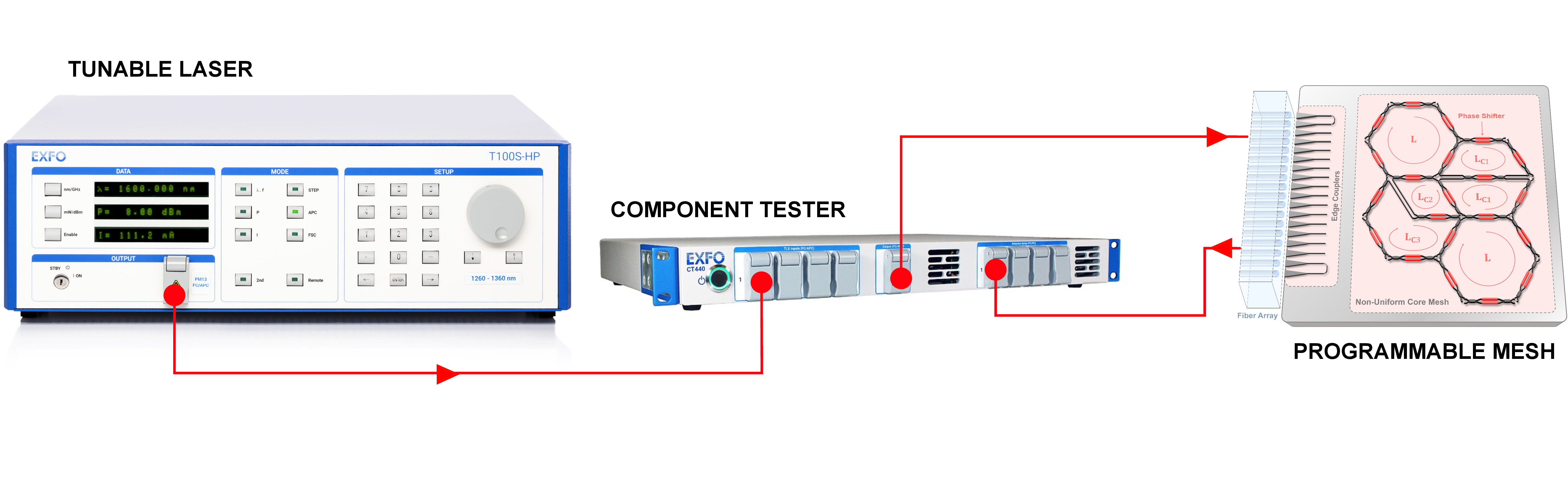}
\caption{Experimental setup for optical spectral measurements, including a tunable laser source, component tester, and programmable photonic mesh.}
\label{fig11s}
\end{figure}

For optical spectral measurements, we used a tunable laser source (T100S-HP EXFO) along with a component tester (CT440 EXFO), featuring a 1~pm wavelength resolution. Figure~\ref{fig11s} shows in more detail the optical connections between the different equipment.
Active optical alignment was performed automatically using piezo controllers (Thorlabs MDT693B) for the positioner that held the fibre array. The alignment routine consists of finding, simultaneously, high optical output power at both loops of the edge coupler array. A 7~dB loss is measured for the loop ports, corresponding to a 3.5 dB loss per input, including the coupling and insertion losses of the edge couplers. 

The 42~DC channels are wirebonded to a printed circuit board (PCB) and controlled with a multichannel current source (Qontrol Ltd.) that can be programmed through coding.

\begin{figure}[h]
\centering
\includegraphics[width=\textwidth]{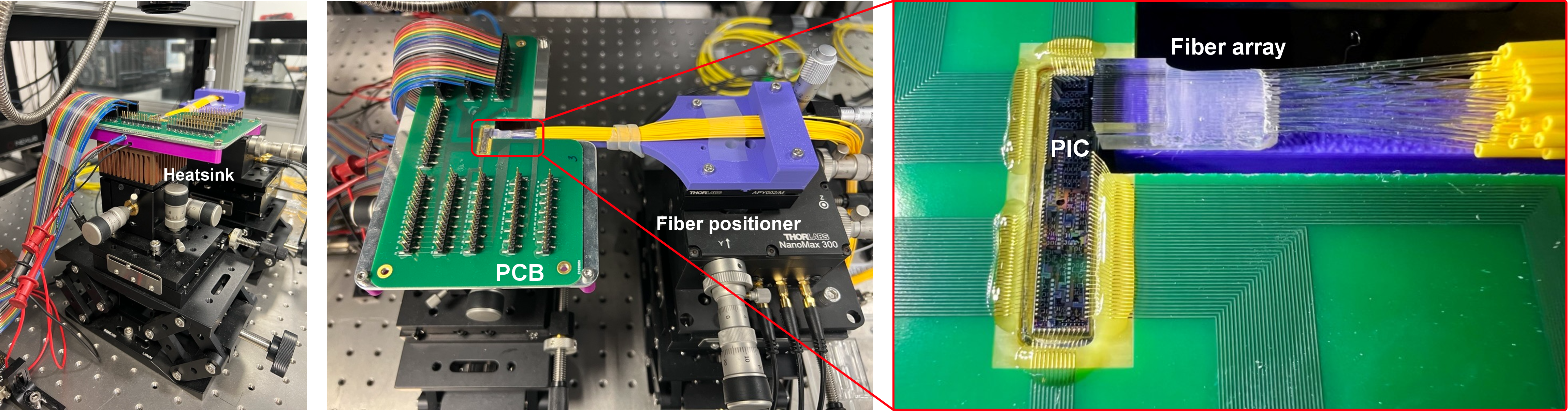}
\caption{Setup for thermal stabilization during measurements, featuring a temperature controller, Peltier cooler, fiber positioner, and printed circuit board (PCB).}
\label{fig12s}
\end{figure}

To have a stable temperature during measurements, we use a temperature controller (Thorlabs TED200C) along with a Peltier cooler and a thermistor. A heatsink was placed under the whole packaged device for heat dissipation as shown in Fig.~\ref{fig12s}, with a more detailed view of all the components.

\newpage
\section*{Supplementary Note 6. \normalfont Characterization of the different cells}\label{sec6}
\begin{figure}[h]
\centering
\includegraphics[width=0.7\textwidth]{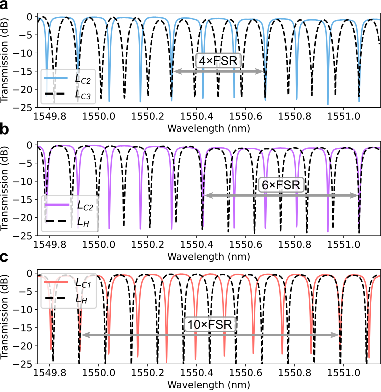}
\caption{Spectral measurements of individual cavities within the programmable mesh, verifying the feasibility of the Vernier effect by comparing resonances of different-sized cavities.}
\label{fig13s}
\end{figure}

Once the entire mesh was characterized, each of the designed cavities was characterized separately by programming simple rings with a random coupling. In Fig.~\ref{fig13s} you can see the spectral measurement of each of these cavities, using the setup in Fig.~\ref{fig11s}.

In Fig.~\ref{fig13s}a, the FSR of the $L_{C2}$ ring is compared with the $L_{C3}$ ring, seeing how their resonances coincide with every 4 FSRs of the $L_{C3}$ cavity. Figure~\ref{fig13s}b,c shows the measurements for the $L_{C2}$ and $L_{C1}$ cavities compared to the hexagonal cavity, showing the resonances that coincide every 6 and 10 FSRs, respectively. Therefore, this is a preliminary way of verifying that it is possible to observe the Vernier effect demonstrated in this work, by coupling each pair of these cavities.

\newpage
\section*{Supplementary Note 7. \normalfont Kramers-Kronig relations for phase response}\label{sec7}
The Kramers-Kronig relations establish a fundamental connection between the real and imaginary parts of a complex function that is analytic in the upper half-plane \cite{ref7}. The fundamental principle behind these relations is causality: the response of a system at a given time depends only on past interactions. This constraint imposes mathematical relationships between different components of a function in the frequency domain, allowing one part to be determined if the other is known.

For a causal response function $H(\omega)$, the real and imaginary components are related via the Hilbert transform:

\begin{equation}
    \operatorname{Re}[H(\omega)] = \frac{1}{\pi} \mathcal{P} \int_{-\infty}^{\infty} \frac{\operatorname{Im}[H(\omega')]}{\omega' - \omega} d\omega',
    \label{eq:KK_real}
\end{equation}

\begin{equation}
    \operatorname{Im}[H(\omega)] = -\frac{1}{\pi} \mathcal{P} \int_{-\infty}^{\infty} \frac{\operatorname{Re}[H(\omega')]}{\omega' - \omega} d\omega',
    \label{eq:KK_imag}
\end{equation}

where $\mathcal{P}$ denotes the Cauchy principal value. These integrals ensure that if one part of the function is known over all frequencies, the other can be uniquely determined, provided the function satisfies the necessary analyticity and decay conditions.

If we consider a complex function of frequency $H(\omega) = A(\omega) e^{i\phi(\omega)}$, where $A(\omega)$ is the amplitude and $\phi(\omega)$ is the phase, the phase can be directly obtained from the amplitude using:

\begin{equation}
    \phi(\omega) = \frac{2\omega}{\pi} \mathcal{P} \int_{0}^{\infty} \frac{\ln A(\omega')}{\omega'^2 - \omega^2} d\omega'.
    \label{eq:KK_phase}
\end{equation}

This relation allows phase reconstruction from measured amplitude spectra. The practical application of these relations requires numerical techniques for evaluating the principal value integral. Common approaches include Fourier-based Hilbert transforms and discrete numerical integration methods.

\newpage
\renewcommand{\refname}{Supplementary References}
 \label{ref}

\end{document}